\documentclass[twocolumn,aps,prl,floatfix,superscriptaddress]{revtex4-2}
\usepackage{amsmath,amssymb}
\usepackage{graphicx,float}
\usepackage{times,txfonts}
\usepackage{color}
\usepackage{multirow}
\usepackage{bbold}
\usepackage{color}
\usepackage{soul}
\usepackage{hyperref}
\usepackage{times,txfonts}
\usepackage{natbib}
\usepackage{nicefrac}
\usepackage{blindtext}
\usepackage[export]{adjustbox}
\usepackage{mathrsfs}
\usepackage{textcomp}
\usepackage{blkarray}
\usepackage{multirow}
\usepackage[normalem]{ulem}
\newcommand{\abs}[1]{\left| #1 \right|}

\newcommand{\bra}[1]{\left\langle #1 \right|}
\newcommand{\ket}[1]{\left| #1 \right\rangle}
\newcommand{\braket}[2]{\left\langle {#1{\left| \vphantom{#1 #2} \right.} #2} \right\rangle}

\renewcommand{\epsilon}{\varepsilon}

\def\VR{\kern-\arraycolsep\strut\vrule &\kern-\arraycolsep}
\def\vr{\kern-\arraycolsep & \kern-\arraycolsep}
\usepackage{sistyle}
\usepackage{soul}
\definecolor{lightblue}{RGB}{185,210,248}
\sethlcolor{lightblue}

\begin{document}

\title{Fast Adaptive Optics for High-Dimensional Quantum Communications in Turbulent Channels}

\author{Lukas Scarfe}
\affiliation{Nexus for Quantum Technologies, University of Ottawa, Ottawa, K1N 6N5, ON, Canada}

\author{Felix Hufnagel}
\affiliation{Nexus for Quantum Technologies, University of Ottawa, Ottawa, K1N 6N5, ON, Canada}

\author{Manuel F. Ferrer-Garcia}
\affiliation{Nexus for Quantum Technologies, University of Ottawa, Ottawa, K1N 6N5, ON, Canada}

\author{Alessio D'Errico}
\affiliation{Nexus for Quantum Technologies, University of Ottawa, Ottawa, K1N 6N5, ON, Canada}

\author{Khabat Heshami}
\affiliation{National Research Council of Canada, 100 Sussex Drive, Ottawa ON Canada, K1A 0R6}
\affiliation{Nexus for Quantum Technologies, University of Ottawa, Ottawa, K1N 6N5, ON, Canada}

\author{Ebrahim Karimi}
\email{ekarimi@uottawa.ca}
\affiliation{Nexus for Quantum Technologies, University of Ottawa, Ottawa, K1N 6N5, ON, Canada}
\affiliation{National Research Council of Canada, 100 Sussex Drive, Ottawa ON Canada, K1A 0R6}

\begin{abstract}
Quantum Key Distribution (QKD) promises a provably secure method to transmit information from one party to another. Free-space QKD allows for this information to be sent over great distances and in places where fibre-based communications cannot be implemented, such as ground-satellite. The primary limiting factor for free-space links is the effect of atmospheric turbulence, which can result in significant error rates and increased losses in QKD channels. Here, we employ the use of a high-speed Adaptive Optics (AO) system to make real-time corrections to the wavefront distortions on spatial modes that are used for high-dimensional QKD in our turbulent channel. First, we demonstrate the effectiveness of the AO system in improving the coupling efficiency of a Gaussian mode that has propagated through turbulence. Through process tomography, we show that our system is capable of significantly reducing the crosstalk of spatial modes in the channel. Finally, we show that employing AO reduces the quantum dit error rate for a high-dimensional orbital angular momentum-based QKD protocol, allowing for secure communication in a channel where it would otherwise be impossible. These results are promising for establishing long-distance free-space QKD systems.
\end{abstract}	

\maketitle

\noindent\textbf{Introduction--} Quantum Key Distribution (QKD) allows two parties to generate a shared secret key between themselves by taking advantage of the properties of quantum systems~\cite{Pirandola:20}. Since the introduction of the first protocol by Bennett and Brassard~\cite{bennett1984quantum}, many QKD protocols have been explored theoretically~\cite{rivest1978method} and experimentally~\cite{bennett:92}. The original implementations relied on encoding schemes using light’s polarisation degree of freedom, constraining the quantum states to a two-dimensional vector space. However, higher-dimensional QKD protocols, employing unbounded photonics degrees of freedom, were suggested to increase information density per carrier~\cite{Bechmann-Pasquinucci2000,ecker2019overcoming}. There are many photonic degrees of freedom in addition to polarisation, which can be used for encoding information, including frequency, vector modes, and time bins~\cite{reimer2014integrated,brecht2015photon,ndagano2017creation,islam:17}. Here, we employ spatial structure of the lights transverse mode through the orbital angular momentum (OAM) which has been studied in diverse settings including free-space~\cite{vallone2014free,mirhosseini:15}, fibre ~\cite{wang2021high,cozzolino2019orbital}, and underwater~\cite{Sit:17,sit:18,bouchard2018underwater,hufnagel:19}. Optical beams carrying OAM are characterized by an azimuthal-dependent phase of $e^{i\ell\phi}$ where $\phi$ is the azimuthal coordinate and $\ell$ is an integer. Because the OAM modes comprise a complete orthonormal basis, they can be used to implement high-dimensional QKD protocols~\cite{mair2001entanglement}. 
\begin{figure*}[!htb]
	\begin{center}
		\includegraphics[width=2.0\columnwidth]{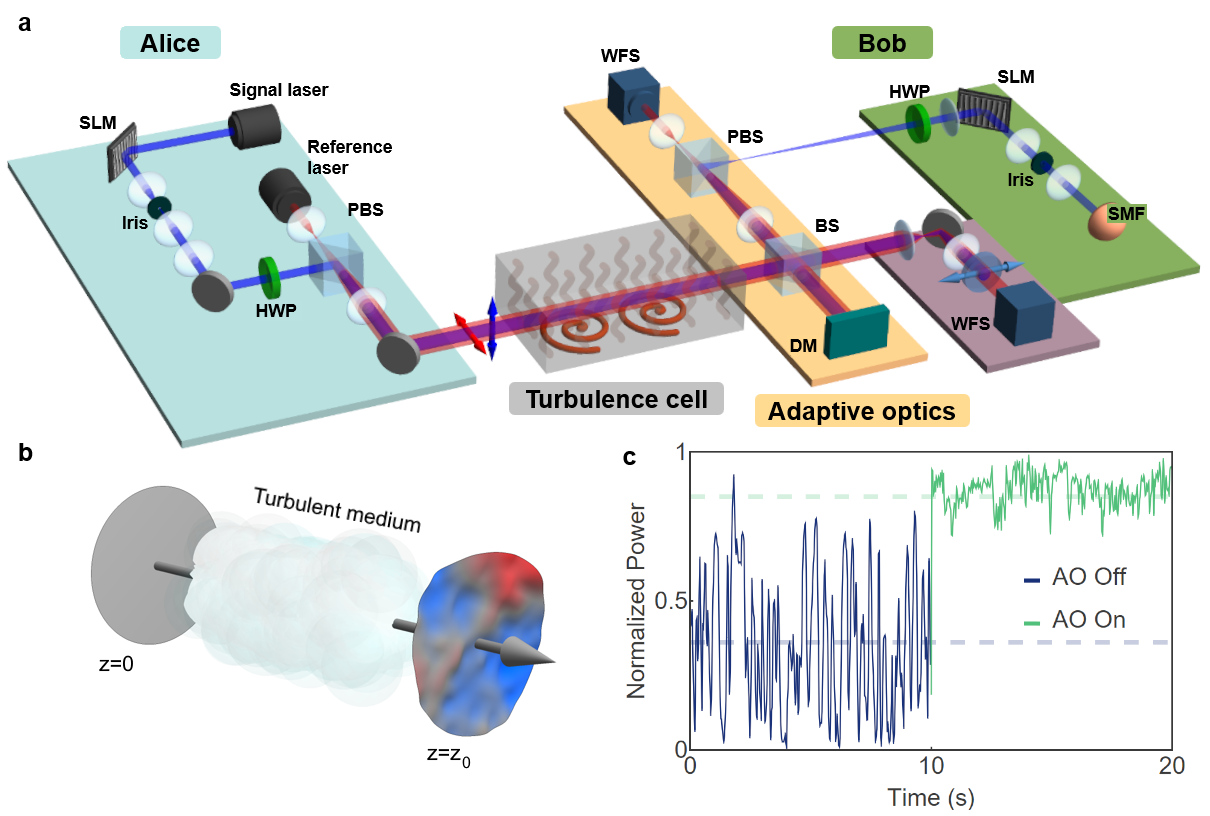}
		\caption{\textbf{High-dimensional quantum communication with adaptive optics through a turbulent channel}. \textbf{a} Experimental setup used to investigate the corrective action of a fast adaptive optics (AO) system (from ALPAO~\cite{alpao}) on structured optical beams after propagation through a turbulent channel. A 633 nm laser impinges on a spatial light modulator (SLM), tailoring the complex field (both amplitude and phase) of the input beam. Additionally, a second laser source of the same wavelength emits vertically polarized light, which is expanded to approximate a plane wave for use as a reference beam. These beams are combined at a polarizing beam splitter (PBS) and sent through a turbulent cell. Here, the turbulence is generated by employing a controllable hotplate placed inside a glass tank with a width of $30$~cm. The composite beam is split using a 50:50 beam splitter; one part goes to a wavefront sensor (WFS) to record the output wavefront, while the second part is fed to the AO section of the experiment. Our AO apparatus consists of a deformable mirror (DM), and a WFS connected in a closed-loop control system. As the WFS measures the structure of the wavefront, the DM changes shape to compensate for the distortions introduced by the turbulence. In our particular experiment, the reference and signal beam are split using a PBS following the corrections applied by DM. Finally, the signal component is sent to a second SLM that performs a projective measurement of spatial modes to determine the probability of detection. \textbf{b} Illustration of the effects on the phase of a plane wave after propagating through a turbulent medium. The colours on the output represent the leading and lagging deformations on the wavefront due to the non-uniform refractive index of the medium. \textbf{c} Normalized optical power coupled into a single mode fibre, as measured by a power meter during the application of turbulence on a Gaussian input beam. The wavefront correction component is activated ten seconds after the beginning of the measurement. A measurement over a longer time interval is depicted in Fig.~\ref{fig:SuppExtendedGaussian} of the Supplementary materials.}
		\label{fig:expset}
	\end{center}
\end{figure*}

The channels most often used to transmit quantum information are fibre and free-space. Optical fibre has the advantage of being a well-developed optical technology with infrastructure that has been built up alongside the increasing reliance on high-speed internet connection. However, in the case of quantum communication, the significant attenuation losses that come with optical fibres creates a fundamental limit on the distance achievable by QKD protocols. This is because quantum signals cannot be amplified in the same way as classical signals; a consequence of quantum no-cloning theorem~\cite{wootters1982single}. In addition, fibre-based solutions rely on an established network, increasing the implementation costs of near-term quantum systems. Despite the significance of fibre-based networks for QKD, it is critical to develop and improve on free-space links for ground-to-ground and ground-to-space quantum communication~\cite{Tobias:07,liao:17,vallone:15}. Space-based quantum communication can help circumvent the distance-rate tradeoff due to exponential loss in fibre-based networks. The successful implementation of QKD over free-space channels depends on the accurate transmission and detection of single photons after propagation through the atmosphere. Rapid changes in the temperature and pressure of the atmosphere result in variations of the refractive index of the air, creating atmospheric turbulence which distorts the beam upon propagation~\cite{kolmogorov1962refinement}. This spatially distributed non-uniform propagation medium induces continuously varying phase aberrations along the optical path of the communication link. It has been shown in previous works that a turbulent environment has a considerable impact, substantially degrading the quantum state, which results in significant errors within the communication channel~\cite{malik2012influence,klug2021orbital,lavery2017free,cox2020structured,Jeongwan:18}. Consequently, the information encoded within the structure of the photons is likely to be lost due to unintended changes in that structure introduced in propagation. In order to implement a realistic high-dimensional free-space QKD system, the system will require compensation for atmospheric turbulence in the channel. One method of correcting distortions in the atmosphere, which is of particular interest, is adaptive optics (AO). While AO has been employed successfully to correct real-time astronomical observations for decades~\cite{beckers1993adaptive,tyson2000introduction,van2004performance}, its potential application for free-space communications has only recently been explored~\cite{majumdar2008free,wang2018performance,liu2016adaptive}.

\begin{figure*}[!htb]
	\begin{center}\includegraphics[width=2\columnwidth]{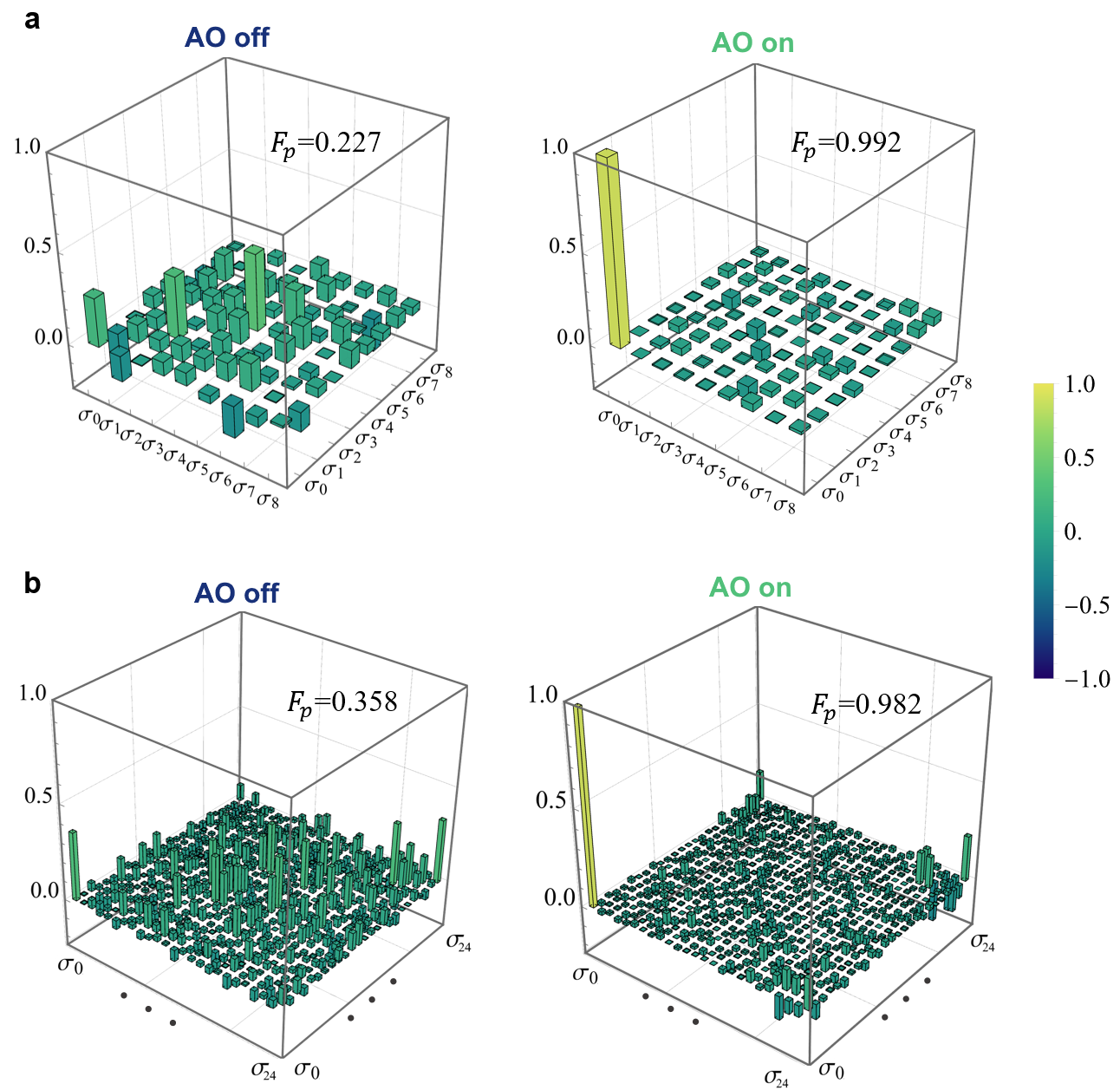}
		\caption{\textbf{Channel Process Tomography for $d=3,5$.} The real part of the process matrix of the channel transmission is shown for dimension 3 with the AO system as (\textbf{a}) non-active and (\textbf{b}) active when going through the turbulence cell. We perform the process tomography using mutually unbiased bases measurements following the methods outlined in~\cite{fernandez2011quantum} (see the Supplemental Material for more information). Fidelity is maintained over 99\% in all cases except for that of turbulence active without adaptive optics (upper right process matrix), where the fidelity falls to 27\%. Here, we see that the effect of the turbulence on the channel is a complete ``channel depolarizing'' of the OAM states, i.e., the existence of huge crosstalk, which is successfully undone with the adaptive optics enabled. The process matrices for dimensions 2, and 4 are provided in the Supplementary Material.}
		\label{fig:process}
	\end{center}
\end{figure*}
In this article, we demonstrate the use of a fast AO system to correct atmospheric disturbances in a free-space quantum key distribution channel when the information is encoded in the photon spatial modes, namely structured photons. First, we show the improved detector coupling efficiency that our AO system is capable of when used to correct the effects of turbulence on a simple Gaussian beam. We then perform quantum process tomography for dimensions two through five under turbulent conditions, both with and without AO active. We calculate the quantum dit error rate (QDER) of the system for even dimensions from 2 through 10 under turbulent conditions, with both AO on and off. We demonstrate a significant improvement in the error rate of the quantum protocol for all dimensions, even in a robust turbulence regime, which results in high crosstalk (high error rates) among the OAM states without AO.

\noindent\textbf{Results}\\
\noindent\textit{Adaptive Optics in the detection stage.--} Let us consider that a free-space channel between Alice and Bob has been deployed, allowing them to exchange information encoded using structured light beams. While propagating, the wavefront is distorted due to its interaction with the atmosphere. To compensate for the effects of the optical turbulence, Bob implements a wavefront-correction stage before decoding the message sent by Alice. A scheme of the proposed experimental setup, which uses an adaptive optics system, is depicted in Fig. ~\ref{fig:expset}\textbf{a}. To take full advantage of the AO system, Alice and Bob use two co-linear (co-propagating) light beams at the same frequency with orthogonal polarisation states. The first component, referred to as the \textit{reference} beam, possesses a Gaussian profile, which has been expanded to approximate a flat wavefront that completely covers our deformable mirror, and also completely overlaps spatially with the second beam, i.e. \textit{signal} beam. This allows us to measure and correct the phase distortions within the channel, either from the optical elements or the environment. The second light beam, the signal beam, serves as our information carrier, where the message is encoded by tailoring its complex amplitude using a spatial light modulator (SLM). It must be noted that since both beams share the optical path, they are subjected to the same atmospheric variations and, therefore, both experience the same distortions. Bob is then capable of correcting the distortion on the signal beam using the phase information obtained from the reference beam. For further details of our experimental implementation, refer to the Supplementary Materials.

As a first step, our signal takes the form of a Gaussian beam. In the presence of optical turbulence and the absence of a correction mechanism, the coupling of the signal to a single-mode fibre at the receiver fluctuates with respect to time due to the wavefront distortion (See Fig. \ref{fig:expset}\textbf{b}). As shown in Fig. \ref{fig:expset}\textbf{c}, when the AO system is inactive, the measured power presents strong fluctuations due to the influence of the introduced turbulence. These effects lead to an average coupling power into the single mode fibre around 36.6\% of the value expected without any turbulence applied. Ten seconds after the beginning of the measurement, the AO system is activated, increasing the average measured power to 87.1\% and stabilizing the coupling efficiency. If this channel were to be used for free-space polarisation QKD, implementing AO improves the coupling efficiency and thus would have resulted in a doubling of the secret key rate. From these results, it is possible to observe the promising benefits of including a fast AO system in the detection stage for many kinds of free-space communications. \\

\noindent\textit{Process Tomography.--} We perform quantum process tomography to determine the effect of the turbulent channel on the OAM states up to $d=5$, i.e., $\ell = \left\{\,-2,-1,0,1,2\,\right\}$. The results show that the channel fidelity deteriorates significantly under the presence of turbulence. Quantum process tomography is used to determine the effect of a process on quantum states~\cite{nielsen2010quantum}. A quantum process $\mathcal{E}$ can be represented using the process matrix $\chi_{mn}$ to describe how input states $\rho_{in}$ are transformed to output states $\rho_{out}$ by
\begin{equation}
    \rho_{out} = \mathcal{E}\,(\rho_{in}) = \sum_{m,n} \chi_{mn} \hat{\sigma}_m\rho \hat{\sigma}_n^{\dagger},
\end{equation}
%\begin{equation}
%        \ket{\psi_0} \ket{\psi_1} \ket{\psi_3} \ket{\psi_5} \ket{\psi_7} \ket{\psi_9} \\
%    \ket{\varphi_0} \ket{\varphi_1} \ket{\varphi_3} \ket{\varphi_5} \ket{\varphi_7} \ket{\varphi_9}
%\end{equation}
\noindent with the Gell-Mann matrices, $\hat{\sigma}_m$ being the high-dimensional extension of the Pauli matrices and satisfying $\sum_{m} \hat{\sigma}^{\dagger}_m \hat{\sigma}_m=\hat{1}$. We seek to determine the process matrix $\chi_{mn}$ by making projective measurements in the high-dimensional mutually unbiased bases (MUB). These projection measurements are described by the operators $\Pi_m^{(\alpha)}$ where the index $\alpha$ denotes the basis and $m$ denotes the state in that basis. It has been proven that for dimensions $d$ that are prime or the power of a prime number, there exists $d+1$ MUBs~\cite{wootters1989optimal}. Thus, in the dimensions explored here, $d=\{2,3,4,5\}$, it is convenient to use the MUB approach to perform process tomography. For an arbitrary dimension, symmetric, informationally complete, positive operator-valued measures (SIC-POVMs) can be used to perform process tomography. The MUB measurement operators in dimension $d$ satisfy
\begin{equation}
\begin{aligned}
    Tr[\Pi_m^{(\alpha)} \Pi_n^{(\alpha)}] = \delta_{mn}, \\
    Tr[\Pi_m^{(\alpha)} \Pi_n^{(\beta)}] = \frac{1}{d},
\end{aligned}
\end{equation}
respectively for the operators of the same basis and different basis, i.e., $\alpha \neq \beta$. Quantum process tomography using MUBs is described in detail in~\cite{fernandez2011quantum}.

\begin{figure}
    \centering
    \includegraphics[width=1.0\columnwidth]{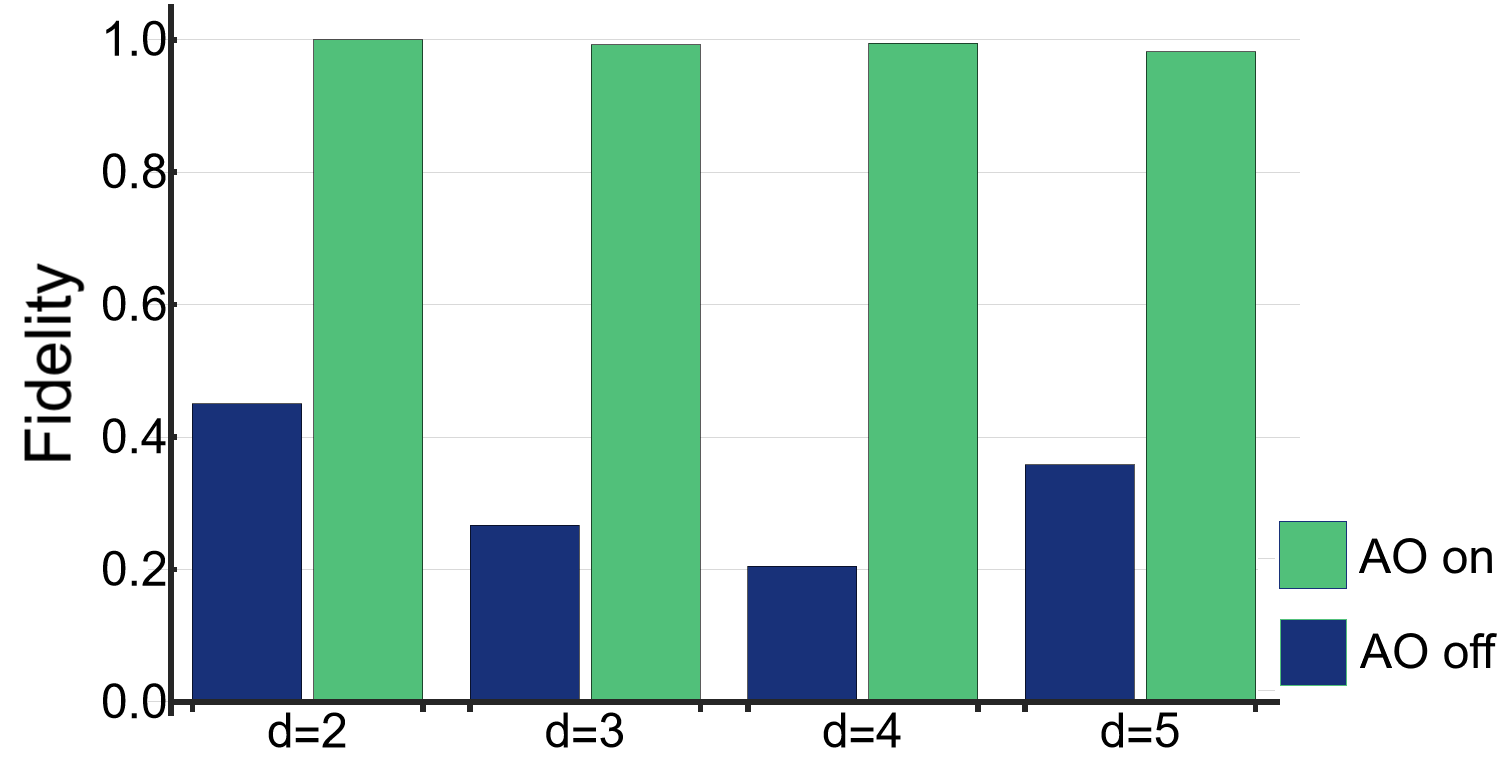}
    \caption{\textbf{Channel Fidelity for OAM-based QKD system.} The process fidelity between the tomographically measured turbulent channel with AO off (blue) and AO on (green) are measured for a QKD channel of $d=2$, $d=3$, $d=4$ and $d=5$. Turbulence `depolarizes' the channel significantly, i.e. introduces huge crosstalk, while activating a fast AO system compensates for the turbulence effects and recovers the encoded states. Due to the long time required to perform these measurements in higher dimensions, the data for $d=5$ was taken on a different day with minor changes to the alignment. This is the main reason why the fidelity for $d=5$ is higher than for $d=3,4$, which were taken one after the other.}
    \label{fig:tomography_bar}
\end{figure}

The channel fidelity for OAM-based QKD without applied turbulence remains high. As the next step, turbulence is applied, and the state tomography is repeated for each dimension. Without any applied turbulence in the channel, in all dimensions, the channel fidelity remains above $\mathcal{F}_p\geq 0.95$. After applying turbulence to the channel and repeating the tomography, the fidelity of the channel is reduced as low as $\mathcal{F}_p\leq 0.45$, indicating a high crosstalk among the modes. The state tomography is repeated with AO enabled in both a turbulent and still environment. The results for the process tomography for $d=3$ are shown in Fig. \ref{fig:process}. In the case of $d=3$, we find that the fidelity of the state is maintained such that $\mathcal{F}_p\geq 0.98$ both with and without turbulence when using adaptive optics. The fidelities of the turbulent channel for all measured dimensions are shown in Fig.~\ref{fig:tomography_bar}. Further process matrices can be found in the Supplementary Material.\\

\begin{figure*}[t]
	\begin{center}
		\includegraphics[width=2\columnwidth]{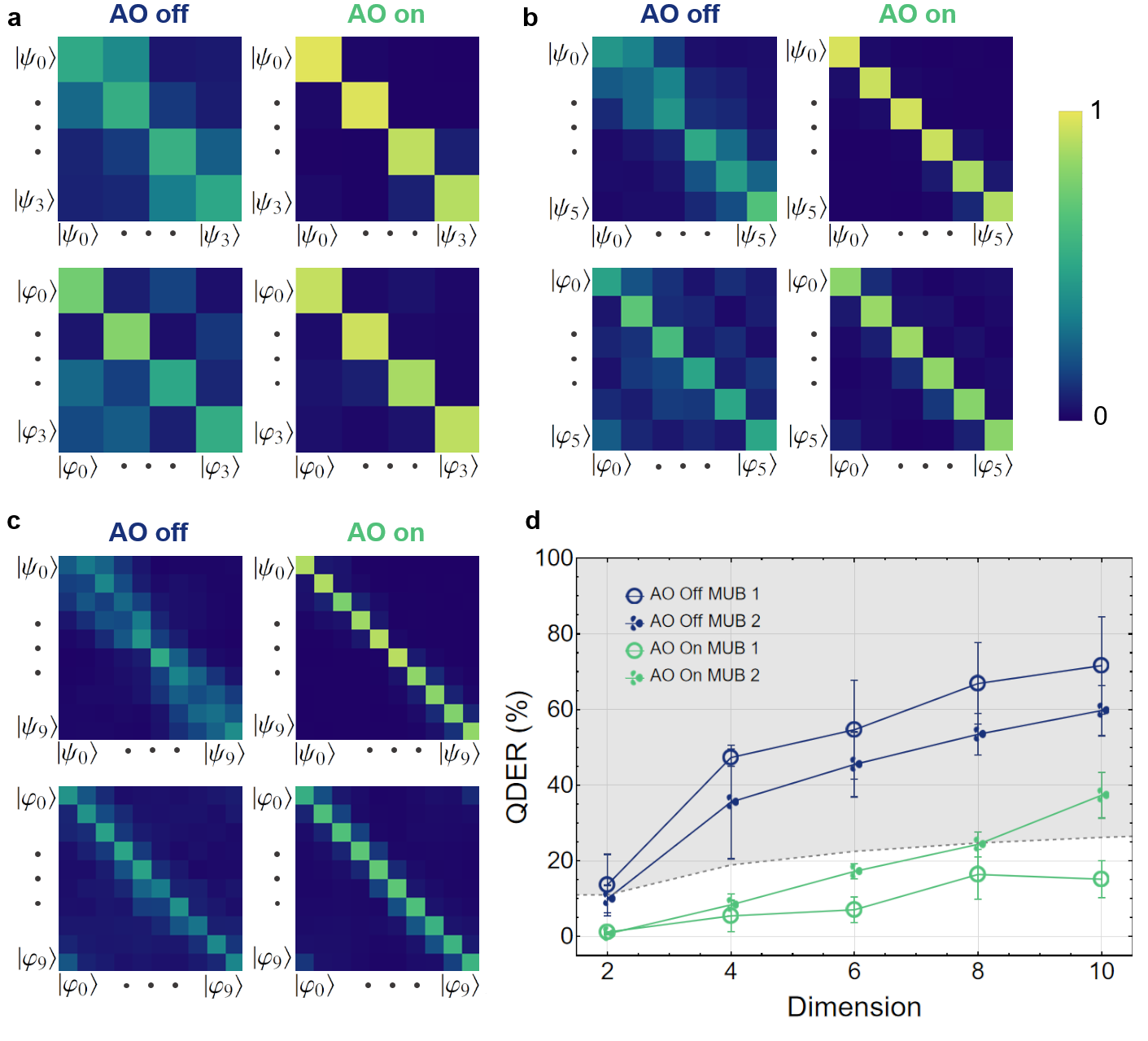}
		\caption{ \textbf{Crosstalk and quantum dit error rate.} The probability of detection on each basis for both bases when going through a turbulent channel for \textbf{a} $d=4$, \textbf{b} $d=6$, and \textbf{c} $d=10$. \textbf{d} Plot of the QDER as calculated from the probability of detection matrices for the cases of adaptive optics on and off with turbulence active. The dashed gray boundary line separates the region for which the theoretical threshold value for QDER allows for a secure key to be established between Alice and Bob. While the turbulent channel prevents communication for any dimension greater than $d=2$ when the correction system is not considered, Bob's use of AO allows for secure keys to be established for all cases less than $d=10$.}
		\label{fig:CrosstalkMatrix}
	\end{center}
\end{figure*}

\noindent\textit{Quantum Dit Error Rate and Crosstalk Matrices.--} To successfully generate a secure key using QKD, it is essential for Bob to accurately detect the state generated by Alice when they choose to operate on the same basis. Any incorrectly detected states will result in a discrepancy between Alice's and Bob's keys, which is quantified as the quantum dit error rate (QDER) $Q$. It must be noted that the maximum value for QDER that is tolerable increases with the dimensionality of the key distribution protocol~\cite{Bechmann-Pasquinucci2000,ecker2019overcoming}. In the case of $d$-dimensional BB84 protocol, the  number of bits of secret key established per sifted photon $R$ is given by~\cite{Bouchard2018experimental},
\begin{equation}
    R(Q) = \text{log}_2(d) -2h(Q),
    \label{eq:Rate}
\end{equation}
where $Q$ is the quantum dit error rate and $h(x)= -x\text{log}_2(x/(d-1))-(1-x)\text{log}_2(1-x)$ is the Shannon entropy. From Eq.~\eqref{eq:Rate}, it is possible to find the QDER threshold when $R=0$. 

Here, the quantum communication channel makes use of two MUB based on two sets of structured beams. The first one, which we consider the logical basis $\{ \ket{\psi_\ell} \}$, is given by the family of OAM states with topological charge $\ell$, where $\ell$ is an integer number. To reduce crosstalk, we consider all values of $\ell = -d/2 \ldots d/2$, excluding the value of $\ell=0$.  Meanwhile, the second MUB, known as the angular mode basis (ANG), consists of a set of beams that are a balanced superposition of such OAM modes given by a quantum Fourier transform of the OAM modes.
\begin{equation}
    \ket{\varphi_k}=\frac{1}{\sqrt{d}}\sum_{j=0}^{d-1}e^{2\pi i\frac{jk}{d}}\ket{j},
\end{equation}

\noindent where $j=d/2+(\ell-1)\,\Theta(\ell)+\ell\,\Theta(-\ell)$, and $\Theta(x)$ is the Heaviside function.

In order to obtain the QDER of a turbulent free-space channel, we need to calculate the crosstalk matrix. A crosstalk matrix is determined by sending each of the states in both bases $\left\{ \ket{\psi_i}\right\}$ and $\left\{ \ket{\varphi_j} \right\}$, and performing projective measurements of the same states. Based on the properties of MUB, it must be noted that a measurement of a projection made on the incorrect basis, i.e. $\abs{\braket{\psi_i}{\varphi_j}}^2$, is equally likely to result in any of the states of the projection basis with a probability of $1/d$. We perform the projective measurements for all even dimensions up to 10, i.e., $d=\{2,4,6,8,10 \}$. The experimental crosstalk matrices in dimensions $d=4$, $d=6$, and $d=10$ for both MUBs in our turbulent channel are shown in panels \textbf{a}, \textbf{b}, and \textbf{c} of Fig. ~\ref{fig:CrosstalkMatrix}, respectively. 

Following these results, we proceed to calculate the QDER of our turbulent channel. Fig. \ref{fig:CrosstalkMatrix}\textbf{d} depicts the QDER as calculated in each dimension $d$. The results show that the QDER exceeds the security boundary given by equation~\eqref{eq:Rate} in all dimensions where $d>2$, measured when no compensation is applied in Bob's detection stage. Therefore, it is not possible to establish a secure communication channel in the presence of applied turbulence. Nevertheless, when the AO system is active, the QDER is reduced to values below the theoretical threshold for positive key rates in all tested cases except for that of the 10-dimensional ANG basis, while. We find that the average decrease in QDER over all tested cases is 32.5\%. This is a promising result, indicating that the use of an AO system can allow for significant improvements in the detection of high-dimensional spatial modes for use in free-space communication.

Our measurements of QDER for different QKD dimensions are shown in Fig. ~\ref{fig:CrosstalkMatrix}\textbf{d}. Interestingly, our results show that the logical basis is more influenced by the introduced turbulence than the ANG basis and the AO performs better on reducing the crosstalk in the logical basis rather than in the ANG basis (this is particularly evident for $d=10$). These effects can be qualitatively  understood when considering that both (mild) turbulence and adaptive optics mainly affect the phase of the beam. The orthogonality of OAM modes depends on their azimuthal phase structure, so is extremely sensitive to phase distortions while ANG modes have a smooth phase dependence but different intensity distributions. The AO performs less well on ANG modes in higher dimensional basis since these modes are increasingly localised in the azimuthal coordinate, thus a higher resolution is needed to compensate for the aberrations induced by turbulence.

%We also find that our AO system is more successful at correcting the effects of turbulence on the logical basis, i.e. OAM bases $\left\{\ket{\ell}\right\}$. As we implement higher dimensions, the QBER of the ANG basis increases faster than that of the logical basis, with a secure key unable to be generated in the case of $d=10$ due to the QBER of the ANG basis being significantly above the security boundary. We believe the limited number of DMDs in our AO system and the absence of cylindrical symmetry in ANG may have contributed to this issue. 
Our experiment demonstrates that the use of a sufficiently advanced adaptive optics system can allow for high-dimensional quantum communications in channels where turbulence would otherwise prevent it.\\

%% MAYBE -> \textcolor{red}{Additionally, using the same equation, we have calculated the theoretical secret key rate of our system for all of the dimensions tested.}

\begin{figure}[!htb]
	\begin{center}
		\includegraphics[width=1 \columnwidth]{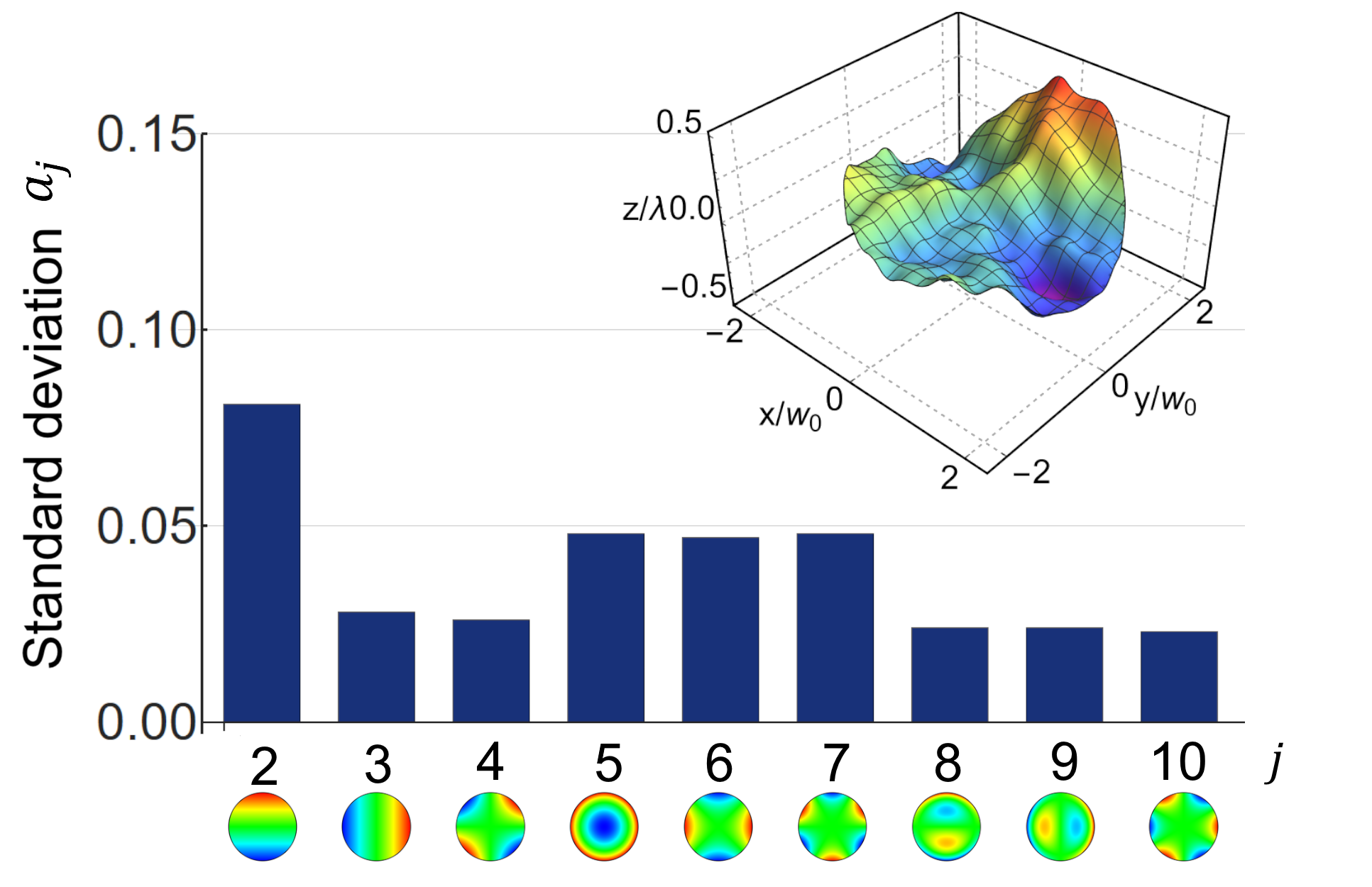}
		\caption{\textbf{Decomposition of the turbulence on the Zernike basis.} Standard deviation of the first nine coefficients $a_j$ of the Zernike decomposition after propagation through the turbulent cell. The WFS reports the turbulence decomposition in the Zernike polynomials basis as a function of time. The inset depicts an aberrated wavefront at a particular time $t_0$, where $w_0$ is the beam waist of the Gaussian mode that would be included in the protocol if using odd dimensions}
		\label{fig:zernikeStats}
	\end{center}
\end{figure}

\noindent\textit{Turbulence Measurement --}
In our experiment, a second WFS is used in our setup in such a way to monitor the reference beam before the correction of the DM was applied (see Fig. ~\ref{fig:expset} \textbf{a}). From the collected data, it is possible to extract instantaneous wavefronts and the corresponding decomposition in terms of Zernike polynomials as functions of time. In Fig. \ref{fig:zernikeStats}, we show standard deviations of the first nine Zernike coefficients, excluding the first one, a global phase shift, over a period of 195 seconds with active turbulence. The strength of the fluctuations in our experiment is in the range of those measured in a previous experiment, where a 3m underwater channel was characterized~\cite{bouchardPool:18}. In this experiment, a secure key could not be generated when $d=4$ due to the effects of the underwater turbulence. In our experiment, we also find that a secure key cannot be generated for $d=4$ unless wavefront correction using a fast AO is implemented in the channel. Thus, our results are promising not only for free-space applications but also for other turbulent environments, i.e. underwater channels.

In addition to measuring the Zernike coefficients, we calculate the Fried parameter $r_0$~\cite{Fried:66,kolmogorov1991local,ageorges2013laser}. This parameter represents the average diameter of the theoretical circular air pockets across which the wavefront phase experiences one radian of variation. From the Fried parameter, we can quantify the strength of the turbulence introduced in our system with the parameter $D/r_0$, where $D$ represents the diameter of the effective aperture used to estimate $r_0$. In our experiment, we obtained $r_0$ by measuring the beam wander of a Gaussian state sent through the channel over time~\cite{Fried:66}. Here, $D$ is given by the waist of the Gaussian beam considered. Following this, we find that the turbulence used in our experiments has a value of $D/r_0=1.70$. This value indicates that our turbulent cell generates moderate-strong turbulence~\cite{zhao2020performance}. This allows us to compare with previous attempts to use active compensation to increase the key rate. Previous studies showed that under similar turbulence conditions, the improvement in QDER when using AO was not enough to establish a secure channel when $d=5$~\cite{zhao2020performance}. This impossibility may result from utilizing an AO system with lower resolution. \\

\textbf{Conclusion --}
In this work, we have tested the capabilities of a fast and high-resolution adaptive optics system in the context of free-space communication channels. We have shown that AO can significantly improve the coupling of a Gaussian beam propagating through a non-uniform, changing medium. Then, we proved the advantage of the use of AO in performing high-dimensional quantum key distribution using spatial modes of photons. Through process tomography, it is shown that the inclusion of the compensation increases fidelity with the identity matrix from under $50\%$ to over $95\%$ in dimensions up to $d=5$. Finally, we demonstrate that by utilizing AO, it is possible to implement a high-dimensional BB84 QKD protocol through a turbulent channel, where it would otherwise not have been possible. We note that the observed turbulence is similar to previously performed experiments both indoors and underwater, as confirmed by Zernike decomposition and the estimation of the Fried parameter. We foresee using an AO system in practical free-space links for classical and quantum communications, in particular, in QKD networks utilizing satellites. 

\vspace{1cm}
\noindent \textbf{Acknowledgments.}
The authors would like to thank Alicia Sit for the valuable discussion and her help in setting up the AO system, as well as the ALPAO support team for their responsibility and support. This work was supported by Canada Research Chairs; Canada First Research Excellence Fund (CFREF); National Research Council of Canada High-Throughput and Secure Networks (HTSN) Challenge Program; and the Qeyssat User INvestigation Team (QUINT) Alliance Consortia Quantum grant.\\

\noindent \textbf{Author Contributions}
E.K. conceived the idea; L.S., F.H., M.F, A.D, and E.K. designed the experiments; L.S. and F.H performed the experiments and collected the data; L.S., F.H., and M.F analysed the data and wrote the first version of the manuscript. K.H. and E.K. supervised the project. All authors discussed the results and contributed to the text of the manuscript.\\

\noindent \textbf{Supplementary materials} accompanies this manuscript. 

%Bibliography
%\bibliographystyle{naturemag}
%\bibliography{AOQKD}

\providecommand{\noopsort}[1]{}

%%%%%%%%%%%%%%%%%%%%%%%%%

\clearpage
\onecolumngrid

\renewcommand{\figurename}{\textbf{Figure}}
\setcounter{figure}{0} \renewcommand{\thefigure}{\textbf{S{\arabic{figure}}}}
\setcounter{table}{0} \renewcommand{\thetable}{S\arabic{table}}
\setcounter{section}{0} \renewcommand{\thesection}{S\arabic{section}}
\setcounter{equation}{0} \renewcommand{\theequation}{S\arabic{equation}}

\begin{center}
{\Large Supplementary Information for: \\ Fast Adaptive Optics for High-Dimensional Quantum \\[1ex] Communications in Turbulent Channels}
\end{center}
%\appendix
\vspace{1 EM}
%\tableofcontents
%\newpage

\section{S1- Turbulence analysis}

\subsection{Optical wavefronts and Zernike polynomials}

\begin{figure*}[b]
	\begin{center}
		\includegraphics[scale=0.4]{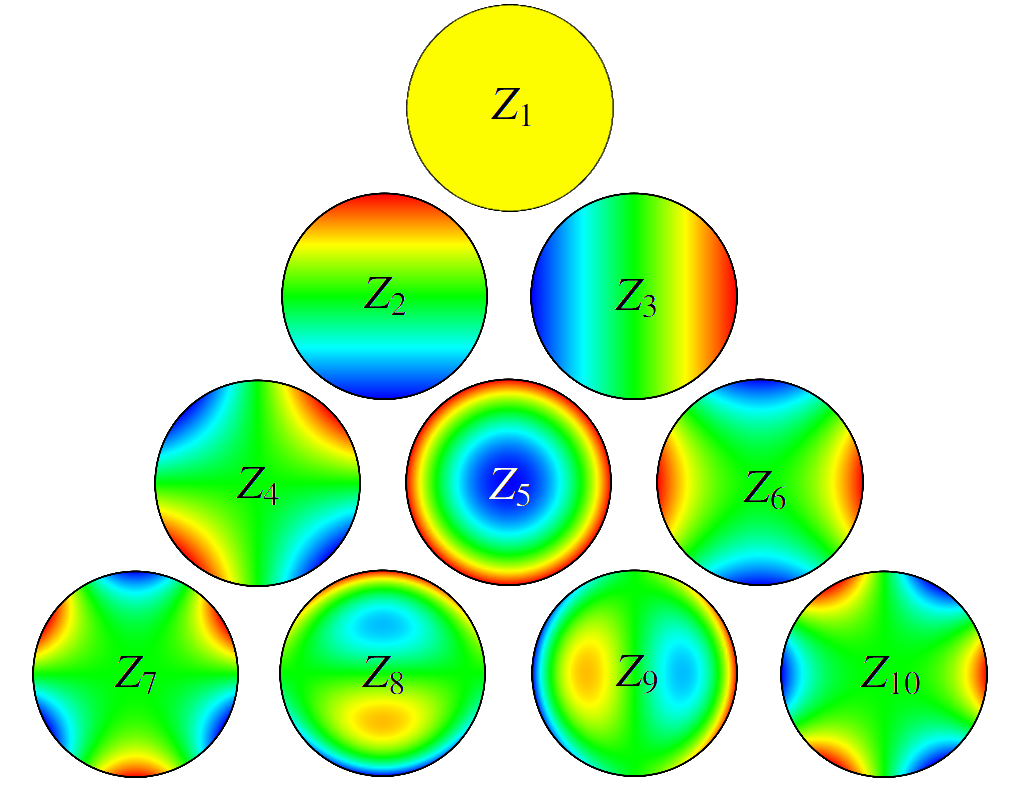}
		\caption{The First 10 Zernike polynomials, ordered vertically by values of $n$ and horizontally by the values of $m$.}
		\label{fig:SuppZernikeModes}
	\end{center}
\end{figure*}

The Zernike Polynomials are a set of orthogonal functions that are defined on a unit circle. Given that the majority of optical systems feature circular apertures, they serve as valuable tools for wavefront analysis and are therefore significant within the field of optics~\cite{born2013principles}. Thus, it is possible to express an arbitrary wavefront $\Phi(R \rho,\phi)$ over a circular aperture of radius $R$ in terms of the Zernike polynomials $Z_j$. Explicitly, we can write
\begin{equation}
\Phi(R \rho,\phi)=\sum_j a_j Z_j(\rho,\phi),
\end{equation}
where $a_j \in \mathbb{R}$ are the coefficients of the expansion and $(\rho,\phi)$ are the cylindrical coordinate system. It must be noted that in this manuscript, we follow the normalized single-index Zernike polynomials according to the ANSI standard~\cite{ANSI}. Table~\ref{tab:Zernike} contains some information regarding the correspondence between the indexes and the Zernike Polynomials. Figure~\ref{fig:SuppZernikeModes} illustrates the first ten Zernike polynomials -- the hue colour shows the function value in the interval of $[-1,+1]$.

\begin{table}[h]
\begin{tabular}{|cc|cc|c|c|}
\hline
\multicolumn{2}{|c|}{ANSI index}         & \multicolumn{2}{c|}{Standard indeces} & {Polynomial} & {Name} \\ \cline{1-4}
\multicolumn{1}{|c|}{Index} & Normalization Factor     & \multicolumn{1}{c|}{ \,\, n  \,\, }     & m      &                             &                       \\ \hline
\multicolumn{1}{|c|}{1}     & 1          & \multicolumn{1}{c|}{0}     & 0      & 1                           & Piston                \\ \hline
\multicolumn{1}{|c|}{2}     & 2          & \multicolumn{1}{c|}{1}     & -1     & $\rho\sin\varphi$            & Tip Y                 \\ \hline
\multicolumn{1}{|c|}{3}     & 2          & \multicolumn{1}{c|}{1}     & 1      & $\rho\cos\varphi$                      & TipX                  \\ \hline
\multicolumn{1}{|c|}{4}     & $\sqrt{6}$ & \multicolumn{1}{c|}{2}     & -2     &  $\rho^2\sin(2\varphi)$        & Astigmatism +45d      \\ \hline
\multicolumn{1}{|c|}{5}     & $\sqrt{3}$ & \multicolumn{1}{c|}{2}     & 0      &  $2\rho^2-1$           & Defocus               \\ \hline
\multicolumn{1}{|c|}{6}     & $\sqrt{6}$ & \multicolumn{1}{c|}{2}     & 2      &  $\rho^2\cos(2\varphi)$      & Astigmatism 0/90d     \\ \hline
\multicolumn{1}{|c|}{7}     & $\sqrt{8}$ & \multicolumn{1}{c|}{3}     & -3     &   $\rho^3\sin(3\varphi)$      & Trefoil Y             \\ \hline
\multicolumn{1}{|c|}{8}     & $\sqrt{8}$ & \multicolumn{1}{c|}{3}     & -1     &   $3\rho^3\sin\varphi-2\rho\sin\varphi$                          & Coma X                \\ \hline
\multicolumn{1}{|c|}{9}     & $\sqrt{8}$ & \multicolumn{1}{c|}{3}     & 1      &  $3\rho^3\cos\varphi-2\rho\cos\varphi$ & Coma Y                \\ \hline
\multicolumn{1}{|c|}{10}    & $\sqrt{8}$ & \multicolumn{1}{c|}{3}     & -3     &  $\rho^3\cos(3\varphi)$ & Trefoil X             \\ \hline
\end{tabular}
\caption{Zernike polynomials are ordered according to their ANSI index, a common alternative indexing scheme, as well as the polynomial in cylindrical coordinates. }
\label{tab:Zernike}
\end{table}

\subsection{Calculation of the Fried Parameter}

Let us define the Fried parameter $r_0$ as a fundamental spatial coherence length measure that quantifies the spatial resolution of the effect of the atmospheric turbulence that our beam experiences. In general, the Fried parameter is given by~\cite{zhan2019wave}  
\begin{equation}
r_0=0.98\frac{\lambda}{\beta},
\end{equation}
where $\lambda$ corresponds to the beam's wavelength while $\beta$ is the average deflection angle experienced by the beam. In our case, the latter is obtained by measuring the position of the centroid of a Gaussian beam after going through the turbulence cell over short intervals of time. Then, it is possible to calculate the average displacement $\bar{s}$ of the beam's centroid from its original position in the absence of turbulence. Finally, the average deflection angle is then given by,

\begin{equation}
\begin{aligned}
\beta=\tan{ \left( \frac{\Bar{s}}{L} \right)},
\end{aligned}
\end{equation}
where $L$ is the length of the turbulent cell.

\begin{figure*}[ht!]
	\begin{center}
		\includegraphics[scale=0.31]{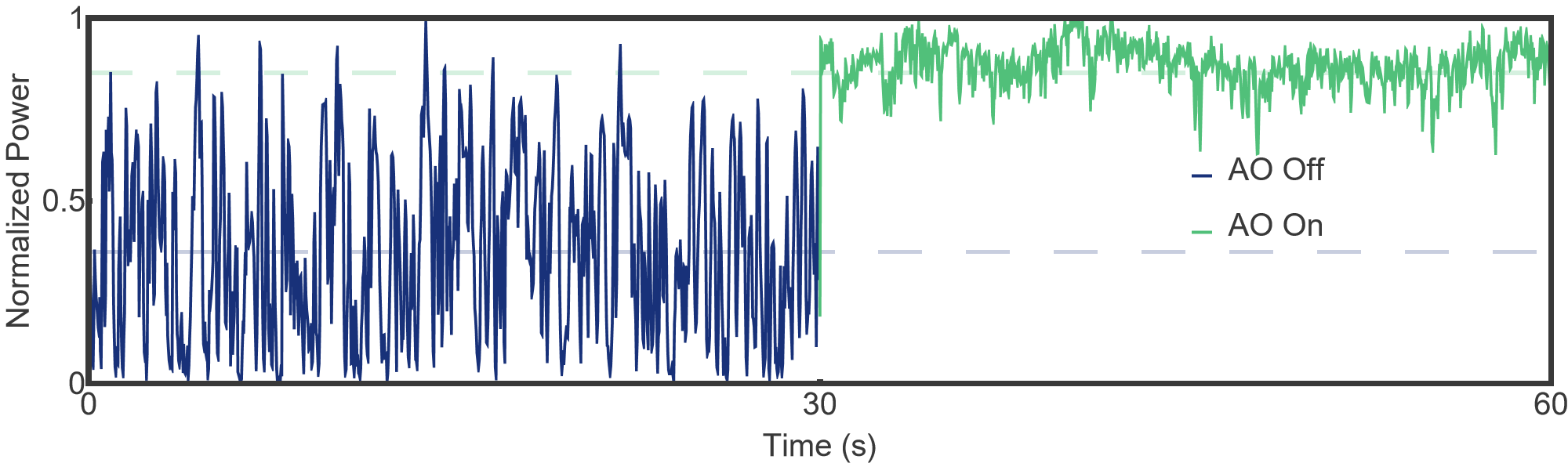}
		\caption{Extended figure showing 60 seconds of a Gaussian beam coupling into a single mode fibre through the active turbulent cell. After 30 seconds without any corrections, the AO system is activated.}
		\label{fig:SuppExtendedGaussian}
	\end{center}
\end{figure*}

\subsection{Adaptive optics system}
The AO system used in our experiment is manufactured by ALPAO and consists of three main components: a deformable mirror (DM) a Shack-Hartmann wavefront sensor (WFS), as well as a feedback-control system. The DM in our configuration (DM9725) has a diameter of $22.5$ mm, and utilizes 97 electromagnetic pistons behind the reflective surface in order to modify its profile. These pistons are organized in an $11\times11$ grid pattern with cut corners to conform to the circular shape of the mirror. It has a settling time of $1.5$ ms, and can therefore operate optimally up to and even slightly above $600$ Hz. On the other hand, the Shack-Hartmannn WFS (SH-EMCCD) has an array of $16\times16$ microlenses in order to correctly measure the reference beam wavefront. It operates at a frequency of 1kHz. The correction calculations are performed by ALPAO Real Time Computer (RTC) and the interface with the whole system is given by using ALPAO Core Engine (ACE) in MATLAB version R2019a Update 3. The system is dependent on the operating frequency to be faster than that of the Greenwood frequency, $f_G$. This frequency is the rate at which the turbulence structure within the optical path changes form \cite{greenwood1977bandwidth}. We can then consider $1/f_G=\tau_G$ to be the Greenwood time constant which is the amount of time that the turbulence structure is constant. During the experiments, the AO system was operating at 200 Hz. While we do not measure the Greenwood frequency of the turbulence generated in the lab, we can be sure that it is less than 200 Hz as the AO system operated without issue. 

\subsection{Extended Gaussian coupling}

\subsection{Turbulent cell}

\begin{figure*}[ht!]
	\begin{center}
		\includegraphics[scale=0.4]{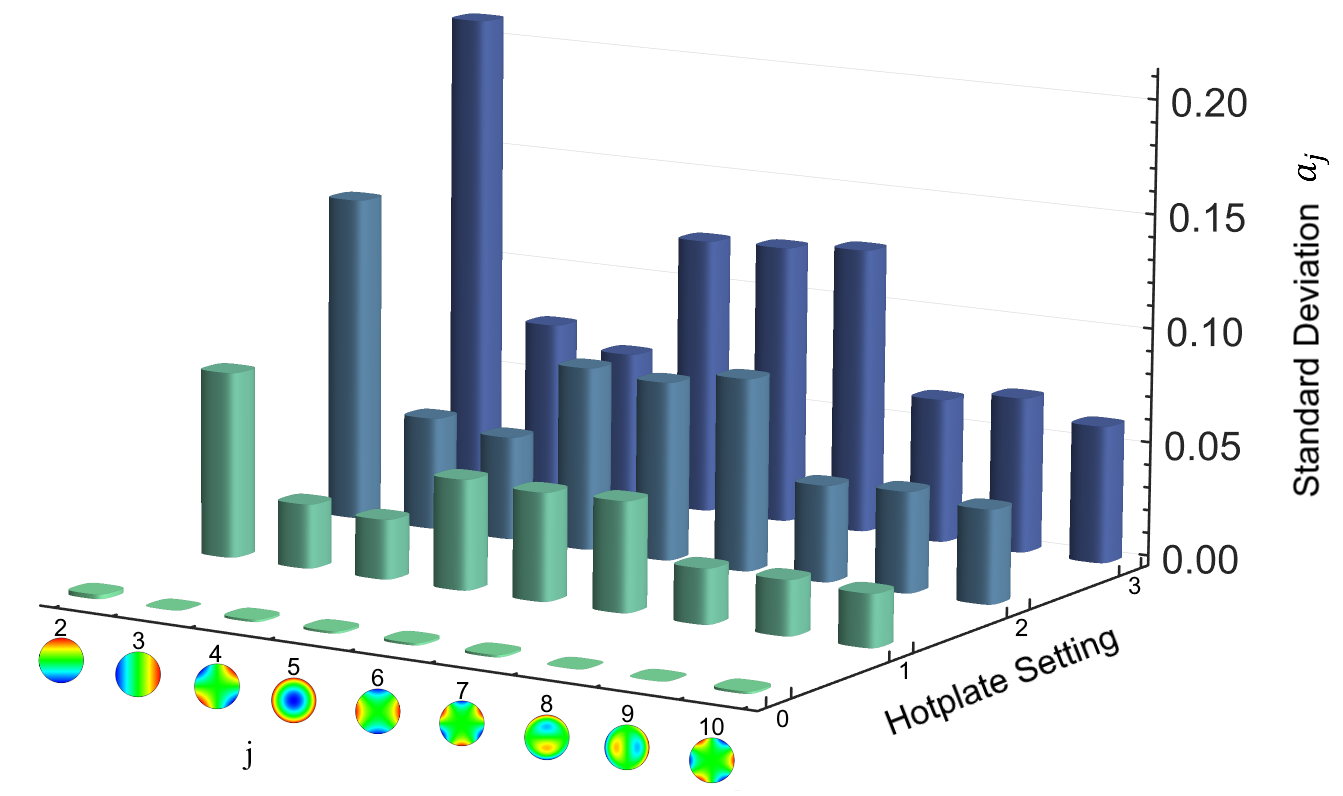}
		\caption{Standard deviations of the first nine coefficients $a_j$ of the Zernike decomposition upon propagation through the turbulent cell as a function of the temperature of the hotplate. Here, the value of 0 corresponds to the hotplate completely off, while 3 stands for the highest temperature possible.}
		\label{fig:Supp3WFS}
	\end{center}
\end{figure*}

In our experiment, the turbulence cell consists of a hotplate contained within a glass-walled water tank. In it, the variations of the refractive index are produced by the temperature gradient generated by the hotplate. As the layer of air close to the plate gets hotter, it rises and displaces the colder layers of air, allowing to generate isolated turbulence inside the tank. The strength of the effective turbulence can be controlled by setting the hotplate at different temperatures. As shown in Fig.~\ref{fig:Supp3WFS}, as the temperature of the hotplate is increased, the standard deviation of the coefficients $a_j$ of the Zernike decomposition also increases. All experiments were performed with the hotplate setting $1$ shown in Fig. \ref{fig:Supp3WFS}.

\section{S2- Process Tomography}
\begin{figure*}[ht!]
	\begin{center}
		\includegraphics[scale=0.37]{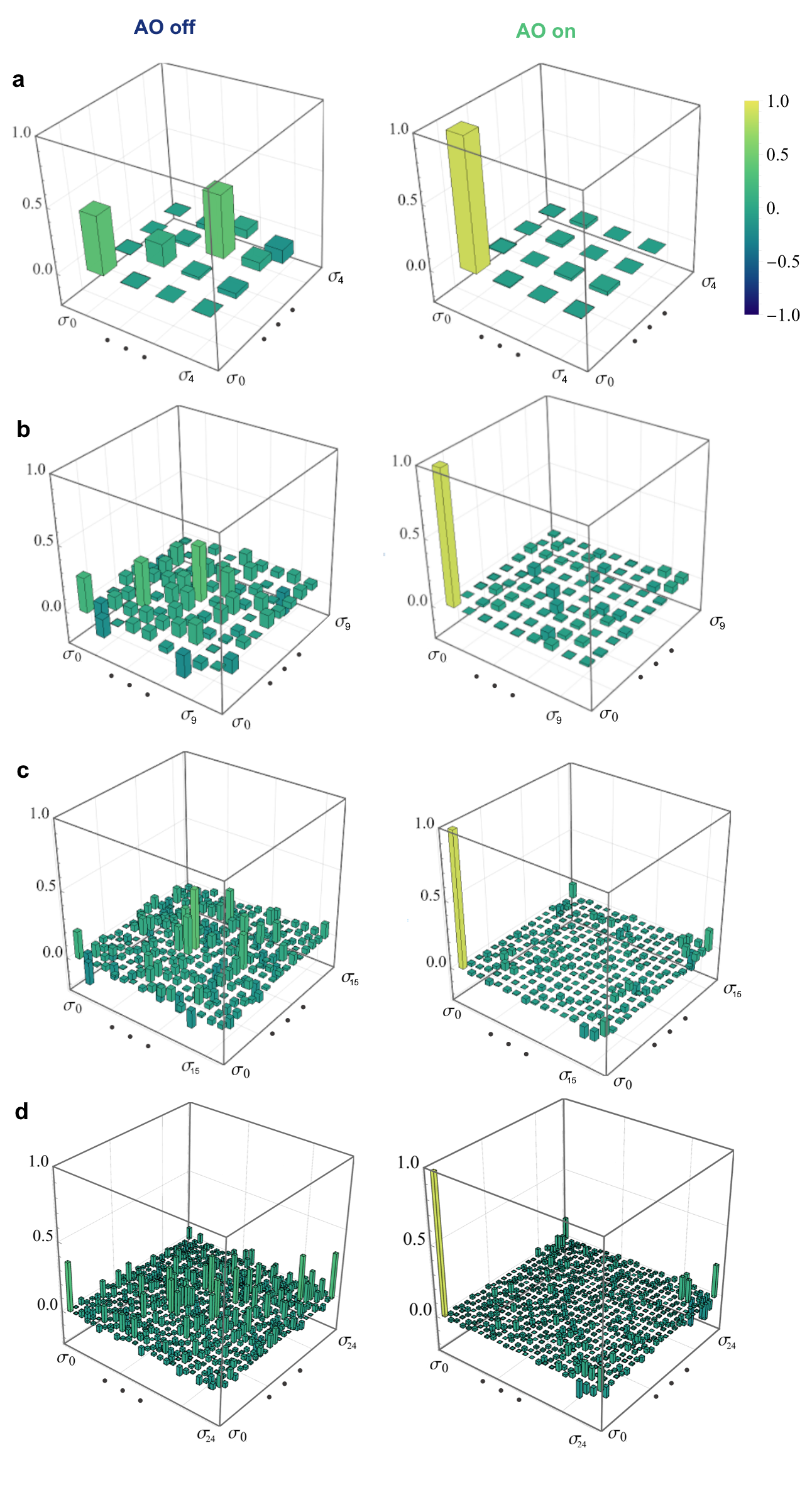}
		\caption{Process matrices for all dimensions. \textbf{a} $d=2$, \textbf{b} $d=3$, \textbf{c} $d=4$, \textbf{d} $d=5$}
		\label{fig:SuppAllTomo}
	\end{center}
\end{figure*}

The process matrices for $d=\{2,3,4,5\}$ are shown in Fig. \ref{fig:SuppAllTomo}. The process tomography was obtained by sending through the turbulent channel and then measuring all the states belonging to the mutually unbiased basis sets for dimension $d$. If $d=p$, where $p$ is a prime numbers, then one can find $p+1$ MUBs. Starting from the canonical basis $\mathcal{B}_0:=\{\ket{j}\}_{j=0\ldots p-1}$, one can generate the basis $\mathcal{B}_{\alpha}:=\{\ket{\psi_0^{\alpha}},\ldots, \ket{\psi_{p-1}^{\alpha}}\}$, with $0\leq {\alpha}\leq p-1$ whose $p$ elements are given by
\begin{align}
\ket{\psi_t^{\alpha}}:=\frac{1}{\sqrt{d}}\sum_{j=0}^{p-1}(\omega^t)^{p-j}(\omega^{-{\alpha}})^{s_j}\ket{j}
\label{MUBexp1}
\end{align}
where $s_j=j+\ldots+(p-1)$ and $\omega=e^{2\pi i/p}$. The process tomography is performed by preparing all the elements of the set $\mathcal{S}:=\{\mathcal{B}_0,\ldots,\mathcal{B}_p\}$ and performing projective measurements on the same set.  Let $\Pi_t^{\alpha}:=\ket{\psi_t^{\alpha}}\bra{\psi_t^{\alpha}}$, the state resulting from the action of the turbulent channel on a basis element is
\begin{align}
    \mathcal{E}(\Pi_t^k)=\sum_{m,n} \chi_{mn}\sigma_m\Pi_t^k\sigma^{\dagger}_n
\end{align}
where $\sigma_m$ are Gell-Mann matrices. 
A measurement in any of the MUBs yields the detection probabilities
\begin{equation}   p^{\alpha,\beta}_{m,n}=\text{Tr}(\Pi_m^{\alpha}\mathcal{E}(\Pi_n^{\beta}))=\sum_{a,b}\chi_{ab}\text{Tr}(\Pi_m^{\alpha}\sigma_a\Pi_n^{\beta}\sigma^{\dagger}_b).
\end{equation}
Through the steps detailed in  Ref. \cite{fernandez2011quantum} the above equation was inverted to find the process matrix $\chi_{mn}$.
The Fidelity between the experimentally reconstructed process matrix $\chi_{exp}$ and a theoretical one $\chi_{th}$ is
\begin{equation}
\mathcal{F}:=\text{Tr}\biggl(\sqrt{\sqrt{\chi_{exp}} \chi_{th}\sqrt{\chi_{exp}}}\biggl)^2.
\end{equation}
In our case $\chi_{th}$ was considered to be the $d$-dimensional identity matrix, corresponding to an ideal channel).

Note that Eq. \ref{MUBexp1} gives a complete set of MUBs for dimensions which are a prime number. For $d$ equal to the power of a prime, complete sets of MUBs can be still found. For $d=4$ one has: 
\begin{align}
\mathcal{B}_0&=\{(1,0,0,0),(0,1,0,0),(0,0,1,0),(0,0,0,1)\}\cr
    \mathcal{B}_1&=\{(1/2,1/2,1/2,1/2),(1/2,-1/2,-1/2,1/2),(1/2,1/2,-1/2,-1/2),(1/2,-1/2,1/2,-1/2)\}\cr
    \mathcal{B}_2&=\{(1/2,i/2,i/2,-1/2),(1/2,-i/2,-i/2,-1/2),(1/2,i/2,-i/2,1/2),(1/2,-i/2,i/2,1/2)\}\cr
    \mathcal{B}_3&=\{(1/2,1/2,-i/2,i/2),(1/2,-1/2,i/2,i/2),(1/2,1/2,i/2,-i/2),(1/2,-1/2,-i/2,-i/2)\}\cr
    \mathcal{B}_4&=\{(1/2,-i/2,1/2,i/2),(1/2,i/2,-1/2,i/2),(1/2,i/2,1/2,-i/2),(1/2,-i/2,-1/2,-i/2)\}.\cr
\end{align}

\section{S3- Crosstalk \& QDER}

\subsection{Bases}

\begin{figure*}[ht!]
	\begin{center}
		\includegraphics[scale=0.25]{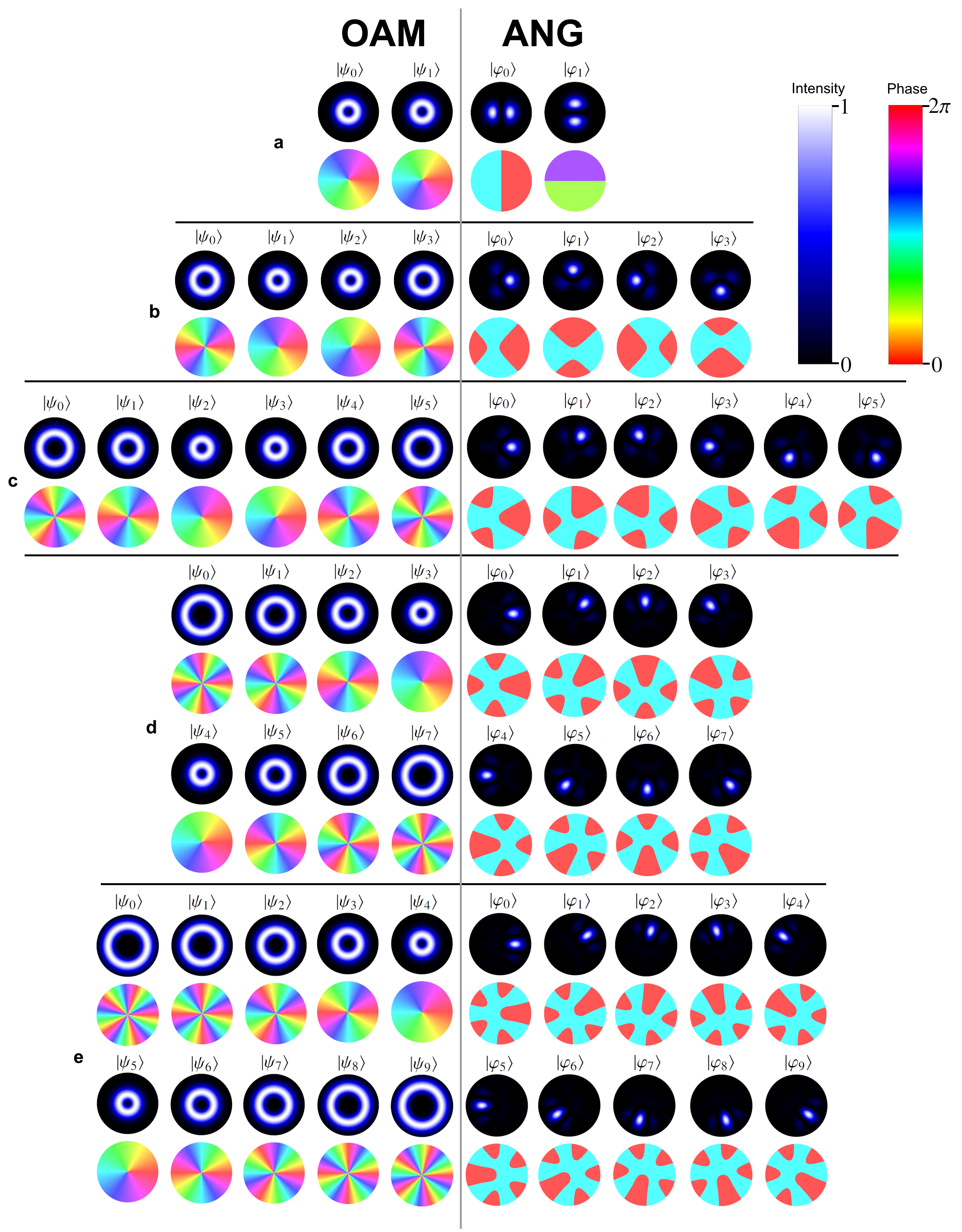}
		\caption{All modes utilized in the crosstalk measurements. The logical basis consisting of OAM modes is shown on the left, while the angular basis is shown on the right for the same dimension. \textbf{a} $d=2$, \textbf{b} $d=4$, \textbf{c} $d=6$, \textbf{d} $d=8$, \textbf{e} $d=10$}
		\label{fig:SuppModes}
	\end{center}
\end{figure*}

As mentioned in the manuscript, we utilize the logical basis, corresponding to OAM modes as out first basis. Our second basis consists of a balanced superposition of the OAM modes corresponding to a quantum Fourier transform known as the angular basis (ANG). The modes for both bases in all dimensions are shown in Fig. \ref{fig:SuppModes}. The phase structure in the ANG basis is made up of flat two regions, with sharp jumps between them. The power in the ANG basis consists of $d$ lobes and becomes more concentrated a single lobe in higher dimensions. 

This localization effectively constrains the mode to fewer 
corrective elements of the adaptive optics system as $d$ increases, not allowing for adequate compensation. This same localization of the ANG states likely allows the state to have a smaller effective diameter, $D$, meaning that the experienced turbulence will be lesser as there is a decrease in $D/r_0$. We believe this is what causes the ANG states to be more robust to turbulence without AO, while also not being as easily corrected when using the AO system. 

\begin{figure*}[ht!]
	\begin{center}
		\includegraphics[scale=0.4]{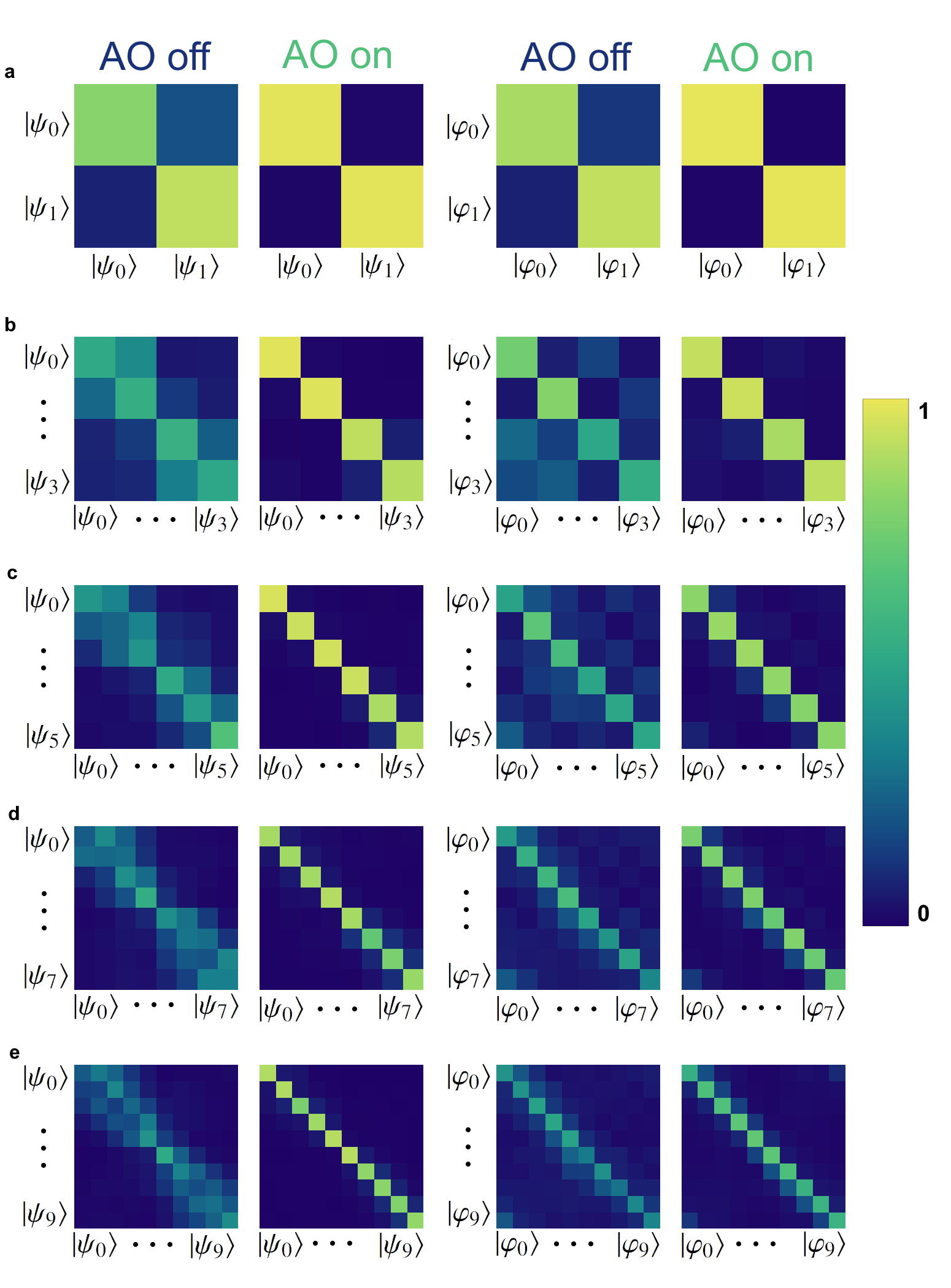}
		\caption{Crosstalk matrices for both bases in all dimensions. \textbf{a} $d=2$, \textbf{b} $d=4$, \textbf{c} $d=6$, \textbf{d} $d=8$, \textbf{e} $d=10$}
		\label{fig:SuppAllCrosstalk}
	\end{center}
\end{figure*}
\subsection{Crosstalk measurement}
For a given basis, the crosstalk matrix is determined through the projective measurement of all states in the basis on the incoming state. After the projective measurement, the light is coupled into a single mode fibre and a power is measured by an optical power meter (Thorlabs PM100D). Each projective measurement $\abs{\braket{\psi_j}{\psi_i}}^2$ is normalized by the total power measured from one incoming state, 

\begin{equation}
\begin{aligned}
C_{ij}=\frac{\abs{\braket{\psi_j}{\psi_i}}^2}{\sum_{j=0}^{d-1}\abs{\braket{\psi_j}{\psi_i}}^2},
\end{aligned}
\end{equation}

\noindent to ensure that the sum of the power measured on an input state (the sum of any row in the matrix) is unitary. This gives the likelihood of detection for any one output state given an input state. These elements are arranged such that Alice's input states are given by the row number, $i$, while Bob's projective measurement state is given by the column number, $j$.

Figure \ref{fig:SuppAllCrosstalk} shows the corresponding crosstalk matrices for all dimensions. We see that the OAM modes are likely to spread to neighboring modes up to the midpoint of the dimension. This shows that the induced turbulence is unlikely to result in power spreading from modes where $\ell >0$ to modes where $\ell<0$ and vice-versa.  

\subsection{QDER calculation}
With the crosstalk matrix measurement performed, the average of the diagonal elements is used to determine the fidelity of the basis. To determine the quantum dit error rate, we subtract the fidelity from the theoretical best performance of 1. This gives a QDER for the basis in a dimension d.

\begin{equation}
\begin{aligned}
\text{QDER}=1-{\sum_{j=i=0}^{d-1}C_{ii}/d}=1-\frac{1}{d}\,\text{Tr}[C]
\end{aligned}
\end{equation}

We calculate the QDER for each of the bases, in each dimension. We find that our AO system is capable of correcting the effects of turbulence in the logical basis for al dimensions. As mentioned in the manuscript, the QDER in the ANG basis is brought below the threshold for secure communications in all cases, save for $d=10$. The exact values for the calculated QDER in all cases are listed in Tables \ref{tab:QBEROAM} and \ref{tab:QBERANG}.

\begin{table}[h]
\begin{tabular}{ |c|c|c|c| } 
 \hline
 Dimension & QDER OAM AO off & QDER OAM AO on & Security Boundary \\ 
 \hline
 2 & $13.6 \pm 8.1$ & $1.2 \pm 0.1$ & $11.0$ \\ \hline
 4 & $47.3 \pm 2.3$ & $5.4 \pm 4.1$ & $18.9$ \\ \hline
 6 & $54.6 \pm 13.1$ & $7.1 \pm 3.4$ & $22.5$ \\ \hline
 8 & $66.9 \pm 10.8$ & $16.4 \pm 6.6$ & $24.7$ \\ \hline
 10 & $71.6 \pm 12.9$  & $15.1 \pm 4.9$ & $26.2$\\ \hline
 \hline
\end{tabular}
\caption{Calculated QDER for the logical basis.}
\label{tab:QBEROAM}
\end{table}

\begin{table}[h]
\begin{tabular}{ |c|c|c|c| } 
 \hline
 Dimension & QDER ANG AO off & QDER ANG AO on & Security Boundary \\ 
 \hline
 2 & $9.9 \pm 3.6$ & $0.6 \pm 0.2$ & $11.0$ \\ \hline
 4 & $35.5 \pm 2.3$ & $ 8.3 \pm 2.9$ & $18.9$ \\ \hline
 6 & $45.5 \pm 8.6$ & $17.2 \pm 2.0$ & $22.5$ \\ \hline
 8 & $53.4 \pm 5.5$ & $24.3 \pm 3.3$ & $24.7$ \\ \hline
 10 & $59.7 \pm 6.6$  & $37.3 \pm 6.0$ & $26.2$\\ \hline
 \hline
\end{tabular}
\caption{Calculated QDER for the angular basis.}
\label{tab:QBERANG}
\end{table}

\end{document}